\documentclass[twocolumn]{IEEEtran}
\usepackage[T1]{fontenc}
\usepackage[latin9]{inputenc}
\usepackage{array}
\usepackage{units}
\usepackage{bm}
\usepackage{multirow}
\usepackage{amsmath}
\usepackage{amsthm}
\usepackage{amssymb}
\usepackage{graphicx}
\usepackage[unicode=true,
 bookmarks=true,bookmarksnumbered=true,bookmarksopen=true,bookmarksopenlevel=1,
 breaklinks=false,pdfborder={0 0 0},pdfborderstyle={},backref=false,colorlinks=false]
 {hyperref}
\hypersetup{pdftitle={Your Title},
 pdfauthor={Your Name},
 pdfpagelayout=OneColumn, pdfnewwindow=true, pdfstartview=XYZ, plainpages=false}

\makeatletter

\providecommand{\tabularnewline}{\\}

\theoremstyle{plain}
\newtheorem{thm}{\protect\theoremname}

\usepackage[caption=false,font=footnotesize]{subfig}
\usepackage{diagbox}
\usepackage{algorithm}
\usepackage{algorithmic}
\usepackage{color}
\usepackage{amsmath}
\usepackage{bm}

\usepackage[top=0.75in, bottom=0.95in, left=0.625in, right=0.625in]{geometry}

\@ifundefined{showcaptionsetup}{}{%
 \PassOptionsToPackage{caption=false}{subfig}}
\usepackage{subfig}
\makeatother

\providecommand{\theoremname}{Theorem}

\begin{document}
\title{Learning Bayes-Optimal Channel Estimation for Holographic MIMO in
Unknown EM Environments}
\author{\author{\IEEEauthorblockN{Wentao Yu$^*$, Hengtao He$^*$, Xianghao Yu$^\dagger$, Shenghui Song$^*$, Jun Zhang$^*$, \textit{Fellow, IEEE}, \\Ross D. Murch$^*$, \textit{Fellow, IEEE}, and Khaled B. Letaief$^*$, {\textit{Fellow, IEEE}}} \\	\IEEEauthorblockA{ 		$^*$Dept. of Electronic and Computer Engineering, Hong Kong University of Science and Technology, Hong Kong\\	$^\dagger$Dept. of Electrical Engineering, City University of Hong Kong, Hong Kong  \\	Email: $^*$\{wyuaq, eehthe, eeshsong, eejzhang, eermurch, eekhaled\}@ust.hk, $^\dagger$alex.yu@cityu.edu.hk}   \thanks{This work was supported in part by the Hong Kong Research Grants Council under Grant No. 16212922, 16209622, and the Areas of Excellence Scheme Grant  No. AoE/E-601/22-R, in part by a grant from the NSFC/RGC Joint Research Scheme sponsored by the Research Grants Council of the Hong Kong SAR, China and National Natural Science Foundation of China (Project No. N\_HKUST656/22), and in part by the National Natural Science Foundation of China for Young Scientists under Grant No. 62301468.} \thanks{An extended journal version of this work is available at \cite{2023Yu-Bayes}. }  } }
\maketitle
\thispagestyle{empty} 
\begin{abstract}
Holographic MIMO (HMIMO) has recently been recognized as a promising
enabler for future 6G systems through the use of an ultra-massive
number of antennas in a compact space to exploit the propagation characteristics
of the electromagnetic (EM) channel. Nevertheless, the promised gain
of HMIMO could not be fully unleashed without an efficient means to
estimate the high-dimensional channel. Bayes-optimal estimators typically
necessitate either a large volume of \textit{supervised} training
samples or \textit{a priori }knowledge of the true channel distribution,
which could hardly be available in practice due to the enormous system
scale and the complicated EM environments. It is thus important to
design a Bayes-optimal estimator for the HMIMO channels in \textit{arbitrary}
\textit{and unknown} EM environments, \textit{free} of any supervision
or priors. This work proposes a \textit{self-supervised} minimum mean-square-error
(MMSE) channel estimation algorithm based on powerful machine learning
tools, i.e., score matching and principal component analysis. The
training stage requires \textit{only} the pilot signals, without knowing
the spatial correlation, the ground-truth channels, or the received
signal-to-noise-ratio. Simulation results will show that, even being
totally \textit{self-supervised}, the proposed algorithm can still
approach the performance of the oracle MMSE method with an extremely
low complexity, making it a competitive candidate in practice. 
\end{abstract}

\begin{IEEEkeywords}
6G, holographic MIMO, channel estimation, score matching, self-supervised
learning, MMSE estimation
\end{IEEEkeywords}

\section{Introduction}

With an ultra-massive number of antennas closely packed in a compact
space, holographic MIMO (HMIMO) is envisioned as a promising next-generation
multi-antenna technology that enables extremely high spectral and
energy efficiency \cite{2020Pizzo}. To exploit the benefits of the
electromagnetic (EM) channel, it is important to acquire accurate
channel state information. However, this is difficult owing to both
the high dimensionality of the channel and its complicated EM characteristics. 

The minimum mean-square-error (MMSE) estimator is able to achieve
the Bayes-optimal performance in terms of MSE. Implementing it requires
either a perfect knowledge of the prior distribution of the channels
\cite{2022Demir,2023An}, or learning such a distribution from a substantial
number of ground-truth channels \cite{2023Yu-JSTSP,2023Yu-AI}, both
of which are difficult, if not impossible, in HMIMO systems owing
to the extremely large number of antennas. Additionally, the computational
complexity of the MMSE estimator, even the linear version (LMMSE),
is extremely high, since it involves computationally-intensive matrix
inversion operations, which consume a significant amount of computational
budget \cite{2020He}. Existing studies proposed various low-complexity
alternatives, but they all come at the cost of an inferior performance
compared with the MMSE estimator. In \cite{2022Demir}, a subspace-based
channel estimation algorithm was proposed, in which the low-rank property
of the HMIMO spatial correlation was exploited without requiring the
full knowledge of the spatial correlation matrix. In \cite{2023Damico},
a discrete Fourier transform (DFT)-based HMIMO channel estimation
algorithm was proposed by approximating the spatial correlation with
a suitable circulant matrix. Nevertheless, such an algorithm was limited
to uniform linear array (ULA)-based HMIMO systems, and cannot be extended
to the more general antenna array geometries. In \cite{2023An}, a
concise tutorial on HMIMO channel modeling and estimation was presented.
Even though the aforementioned estimators significantly outperform
the conventional least squares (LS) scheme, there still exists quite
a large gap from that of the MMSE estimator. 

In this paper, we affirmatively answer a fundamental question: \textit{Is
it possible to establish a Bayes-optimal MMSE channel estimator for
HMIMO systems in arbitrarily unknown EM environments?} Owing to the
complicated channel distribution and the ultra-high dimensionality
of the problem, classical analytical methods become either sub-optimal
in performance or too complicated to implement. Supervised deep learning-based
methods can achieve near-optimal performance, but highly rely on a
substantial dataset of the ground-truth channels \cite{2023Yu-JSTSP},
which is difficult to achieve in HMIMO systems. The data availability
and complexity constitute the two core challenges that prevents the
practical implementation of the MMSE channel estimator in HMIMO systems.
These challenges can be both tackled by our proposed learning-based
estimator: 
\begin{enumerate}
\item \textit{Data availability}: The proposed estimator needs neither the
prior distribution nor the ground-truth channel data. Only the received
pilot signals are required at the model training stage. 
\item \textit{Complexity}: The proposed estimator drops the prohibitive
matrix inversion, and is with extremely low complexity. 
\end{enumerate}
Specifically, we propose a \textit{self-supervised} deep learning
framework for realizing the Bayes-optimal MMSE channel estimator for
HMIMO. We first prove in theory that the MMSE channel estimator could
be constructed based \textit{solely} on the distribution of the received
pilot signals through the \textit{Stein's score function}. Afterwards,
we propose a practical algorithm to train neural networks to estimate
the score function solely by using the collected received pilot signals.
Lastly, a low-complexity principal component analysis (PCA)-based
method is proposed to estimate the received signal-to-noise-ratio
(SNR) from pilots alone, since it is required in the score-based MMSE
estimator. Simulation results in both isotropic and non-isotropic
environments are provided to illustrate the effectiveness and efficiency
of the proposed estimator. Notably, it achieves almost the same performance
as the oracle MMSE estimator with more than 20 times reduction in
complexity, in a nominal HMIMO setup. 

\textit{Notation:} $a$ is a scalar. $\|\mathbf{a}\|$ is the $\ell_{2}$-norm
of a vector $\mathbf{a}$. $\mathbf{A}^{T}$, $\mathbf{A}^{H}$, $\text{\ensuremath{\Re}}(\mathbf{A})$,
$\ensuremath{\Im}(\mathbf{A})$ are the transpose, Hermitian, the
real part, and the imaginary part of a matrix $\mathbf{A}$, respectively.
$\mathcal{CN}(\boldsymbol{\mu},\mathbf{R})$ and $\mathcal{N}(\boldsymbol{\mu},\mathbf{R})$
are complex and real Gaussian distributions with mean $\boldsymbol{\mu}$
and covariance $\mathbf{R}$, respectively. $\mathbf{I}$ is an identity
matrix with an appropriate shape. 

\section{HMIMO System and Channel Models}

\begin{figure}[t]
\centering{}\includegraphics[width=5cm]{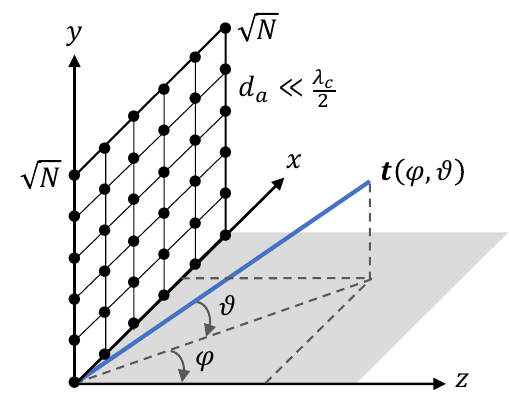}\caption{A UPA-shaped HMIMO BS in a 3D spherical coordinate system with an
impinging plane wave from azimuth AoA $\varphi$ and elevation AoA
$\theta$. \label{fig:system-model}}
\end{figure}

Consider the uplink of an HMIMO system, where the base station (BS)
is equipped with a uniform planar array (UPA) with $\sqrt{N}\times\sqrt{N}$
antennas that simultaneously serves $K$ single antenna-user equipments
(UEs). We focus on the cases where the BS has thousands of closely
packed antennas with spacings $d_{a}$ below the nominal value of
half the carrier wavelength $\lambda_{c}$. We define a local spherical
coordinate system at the UPA with $\varphi\in[-\frac{\pi}{2},\frac{\pi}{2}]$
and $\text{\ensuremath{\vartheta}}\in[-\frac{\pi}{2},\frac{\pi}{2}]$
being the azimuth and elevation angles of arrival (AoAs), respectively,
as depicted in Fig. \ref{fig:system-model}. We index the antennas
row-by-row with $n\in\{1,2,\ldots,N\}$, and denote the position of
the $n$-th antenna as $\mathbf{u}_{n}=[u_{x,n},u_{y,n},0]^{T}$,
in which 
\begin{equation}
\begin{cases}
u_{x,n}=-\frac{(\sqrt{N}-1)d_{a}}{2}+d_{a}\text{mod}(n-1,\sqrt{N}),\\
u_{y,n}=\frac{(\sqrt{N}-1)d_{a}}{2}-d_{a}\lfloor\frac{n-1}{\sqrt{N}}\rfloor.
\end{cases}
\end{equation}
The notations $\text{mod}(\cdot,\cdot)$ and $\lfloor\cdot\rfloor$
refer to the modulus operation and the floor function, respectively.
Considering a planar wave impinging on the UPA\footnote{While we focus on the far-field case here, our proposal is also applicable
to the near-field case \cite{2023Wang}, which will be discussed in
our follow-up works. }, the array response vector is given by 
\begin{equation}
\mathbf{a}(\varphi,\vartheta)=[e^{j\frac{2\pi}{\lambda_{c}}\mathbf{t}(\varphi,\vartheta)^{T}\mathbf{u}_{1}},\ldots,e^{j\frac{2\pi}{\lambda_{c}}\mathbf{t}(\varphi,\vartheta)^{T}\mathbf{u}_{N}}]^{T},
\end{equation}
with $\mathbf{t}(\varphi,\vartheta)=[\cos(\vartheta)\cos(\varphi),\cos(\vartheta)\sin(\varphi),\sin(\vartheta)]^{T}$
being the unit vector in the AoA direction. We assume that orthogonal
pilots are adopted and consider the channel $\mathbf{\bar{h}}\in\mathbb{C}^{N\times1}$
between the BS and an arbitrary UE, consisting of the superposition
of multi-path components that can be represented by a continuum of
planar waves \cite{2022Demir}, given by
\begin{equation}
\mathbf{\bar{h}}=\iint_{-\nicefrac{\text{\ensuremath{\pi}}}{2}}^{\nicefrac{\pi}{2}}g(\varphi,\vartheta)\mathbf{a}(\varphi,\vartheta)d\vartheta d\varphi,
\end{equation}
where $g(\varphi,\vartheta)$ denotes the angular spread function
specifying the phase shift and the gain for each AoA direction $(\varphi,\vartheta)$.
In accordance with \cite{2022Demir}, we can model $g(\varphi,\vartheta)$
as a spatially uncorrelated symmetric Gaussian stochastic process
with cross-correlation given by
\begin{equation}
\mathbb{E}\{g(\varphi,\vartheta)g^{*}(\varphi',\vartheta')\}=\beta f(\varphi,\vartheta)\delta(\varphi-\varphi')\delta(\vartheta-\vartheta'),\label{eq:cross-correlation}
\end{equation}
where $\beta$ is the average channel gain, $\delta(\cdot)$ denotes
the Dirac delta function, and $f(\varphi,\vartheta)$ is the spatial
scattering function, i.e., the joint probability density function
(PDF) of the azimuth and elevation AoAs. The function $f(\varphi,\vartheta)$
is normalized such that $\iint f(\varphi,\vartheta)d\vartheta d\varphi=1$.
The HMIMO channel can be modeled by the correlated Rayleigh fading\footnote{Notice that the proposed algorithm can work with \textit{any possible}
distribution of the HMIMO channel. We follow the convention in the
literature and adopt the correlated Rayleigh fading here. This is
because in this case, the Bayes-optimal estimator admits a closed
form if the covariance $\mathbf{R}$ is perfectly known, which can
serve as the oracle performance bound to benchmark our algorithm. }, i.e., $\mathbf{\bar{h}}\sim\mathcal{CN}(\mathbf{0},\mathbf{R})$,
with a spatial correlation matrix $\mathbf{R}\in\mathbb{C}^{N\times N}$,
which satisfies $\text{tr}(\mathbf{R})=\beta N$ \cite{2022Demir,2023Damico}.
Following (\ref{eq:cross-correlation}), the correlation matrix $\mathbf{R}$
can be calculated by 
\begin{equation}
\mathbf{R}=\mathbb{E}\{\mathbf{\bar{h}}\mathbf{\bar{h}}^{H}\}=\beta\iint_{-\nicefrac{\text{\ensuremath{\pi}}}{2}}^{\nicefrac{\pi}{2}}f(\varphi,\vartheta)\mathbf{a}(\varphi,\vartheta)\mathbf{a}^{H}(\varphi,\vartheta)d\vartheta d\varphi.
\end{equation}
For any function $f(\varphi,\vartheta)$, the $(l,m)$-th entry of
the correlation matrix $\mathbf{R}$ is given by 
\begin{equation}
[\mathbf{R}]_{l,m}=\beta\iint_{-\nicefrac{\text{\ensuremath{\pi}}}{2}}^{\nicefrac{\pi}{2}}e^{j\frac{2\pi}{\lambda_{c}}\mathbf{t}(\varphi,\vartheta)^{T}(\mathbf{u}_{l}-\mathbf{u}_{m})}f(\varphi,\vartheta)d\vartheta d\varphi,\label{eq:covariance-element}
\end{equation}
where $[\cdot]_{l,m}$ denotes the $(l,m)$-th element of a matrix.
While in most cases the integral in (\ref{eq:covariance-element})
could only be computed numerically, a closed form solution exists
in \textit{isotropic} scattering environments, where the covariance
$\mathbf{R}^{\text{iso}}$ is given in \cite{2022Andrea} as
\begin{equation}
[\mathbf{R}^{\text{iso}}]_{l,m}=\text{sinc}(\frac{2\|\mathbf{u}_{l}-\mathbf{u}_{m}\|}{\lambda_{c}}).\label{eq:isotropic-covariance}
\end{equation}
Here $\text{sinc}(\cdot)\triangleq\frac{\sin(\pi x)}{\pi x}$ denotes
the sinc function, while $\|\cdot\|$ is the Euclidean norm. In \textit{non-isotropic}
scattering environments where the augular density is not evenly distributed
in the whole space, $f(\varphi,\vartheta)$ takes many distinct forms
in the literature. For example, in \cite{2022Demir}, the non-isotropic
scattering function $f(\varphi,\vartheta)$ is assumed to follow a
cosine directivity pattern, while in \cite{2023An}, the leading eigenvalues
of $\mathbf{R}^{\text{iso}}$ are truncated to obtain a non-isotropic
covariance, which will be detailed in Section \ref{sec:Simulation-Results}. 

We then introduce the system model. In the uplink channel estimation
phase, the UEs send known pilot sequences to the BS. Assuming that
the orthogonal pilots are utilized, the \textit{real-valued equivalent}
of the received pilot signal (measurement) from an arbitrary UE, $\mathbf{y}\in\mathbb{R}^{2N\times1}$,
is given by\footnote{Similar to \cite{2022Demir}, we consider a fully-digital system model,
in which the dimensions of $\mathbf{y}$ and $\mathbf{h}$ are identical.
Nevertheless, the proposed algorithm can be readily extended to compressed
sensing-based channel estimation in hybrid analog-digital systems,
in a similar manner as \cite{2016Metzler}. }
\begin{equation}
\mathbf{y}=\sqrt{\rho}\mathbf{h}+\mathbf{n},\label{eq:system-model}
\end{equation}
where $\mathbf{h}=[\Re(\mathbf{\bar{h}})^{T},\Im(\mathbf{\bar{h}})^{T}]^{T}\in\mathbb{R}^{2N\times1}$
represents the real-valued channel, $\text{\ensuremath{\rho}}$ is
the received signal-to-noise-ratio (SNR), and $\mathbf{n}\sim\mathcal{N}(\mathbf{0},\mathbf{I})$
is the additive white Gaussian noise. The channel estimator, in practice,
should only have the knowledge of $\mathbf{y}$, without knowing the
true covariance $\mathbf{R}$ or the SNR $\rho$. 

\section{MMSE Estimation via Score Matching}

In the following, we first discuss how to derive the score-based MMSE
estimator solely based on the received pilots $\mathbf{y}$, and then
introduce how to estimate the two key components of the algorithm,
i.e., the score function and the received SNR. 

\subsection{Bridging MMSE Estimation with the Score Function}

Our target is a Bayes-optimal channel estimator, $\mathbf{\hat{h}}=D(\mathbf{y})$,
that minimizes the mean-square-error (MSE), i.e.,
\begin{equation}
\text{MSE}\triangleq\text{\ensuremath{\mathbb{E}(\|\mathbf{h}-\hat{\mathbf{h}}\|^{2}|\mathbf{y})=\int\|\mathbf{h}-\mathbf{\hat{h}}\|^{2}p(\mathbf{h}|\mathbf{y})d\mathbf{h}}},
\end{equation}
where the expectation above is taken with respect to (w.r.t.) the
unknown channel $\mathbf{h}$, while $p(\mathbf{h}|\mathbf{y})$ is
the posterior density. Taking a derivative of the above equation w.r.t.
$\hat{\mathbf{h}}$ and nulling it, we reach the Bayes-optimal, i.e.,
minimum MSE (MMSE), channel estimator, given by
\begin{equation}
\begin{aligned}\hat{\mathbf{h}}_{\text{MMSE}} & =\mathbb{E}(\mathbf{h}|\mathbf{y})=\int\mathbf{h}p(\mathbf{h}|\mathbf{y})d\mathbf{h}=\int\mathbf{h}\frac{p(\mathbf{h},\mathbf{y})}{p(\mathbf{y})}d\mathbf{h},\end{aligned}
\label{eq:posterior-mean}
\end{equation}
where $p(\mathbf{h},\mathbf{y})$ is the joint density, and $p(\mathbf{y})$
is the measurement density obtained via marginalization, i.e., 
\begin{equation}
\begin{aligned}p(\mathbf{y}) & =\int p(\mathbf{h},\mathbf{y})d\mathbf{h}=\int p(\mathbf{y}|\mathbf{h})p(\mathbf{h})d\mathbf{h}\\
 & =\left(\frac{\rho}{2\pi}\right)^{N}\int\exp\left\{ -\frac{\rho}{2}\|\mathbf{y}-\mathbf{h}\|^{2}\right\} p(\mathbf{h})d\mathbf{h}.
\end{aligned}
\label{eq:measurement-density}
\end{equation}

The last equality holds since the likelihood function $p(\mathbf{y}|\mathbf{h})\sim\mathcal{N}(\mathbf{h},\frac{1}{\rho}\mathbf{I})$
expresses $p(\mathbf{y})$ as a convolution between the prior distribution
$p(\mathbf{h})$ and the i.i.d. Gaussian noise. Taking the derivative
of both sides of (\ref{eq:measurement-density}) w.r.t. $\mathbf{y}$
gives 
\begin{equation}
\begin{aligned}\nabla_{\mathbf{y}}p(\mathbf{y}) & =\left(\frac{\rho}{2\pi}\right)^{N}\int\nabla_{\mathbf{y}}\exp\left\{ -\frac{\rho}{2}\|\mathbf{y}-\mathbf{h}\|^{2}\right\} p(\mathbf{h})d\mathbf{h}\\
 & =\rho\left(\frac{\rho}{2\pi}\right)^{N}\int(\mathbf{h}-\mathbf{y})\exp\left\{ -\frac{\rho}{2}\|\mathbf{y}-\mathbf{h}\|^{2}\right\} p(\mathbf{h})d\mathbf{h}\\
 & =\rho\int(\mathbf{h}-\mathbf{y})p(\mathbf{y}|\mathbf{h})p(\mathbf{h})d\mathbf{h}.
\end{aligned}
\label{eq:gradient-py}
\end{equation}
Dividing both sides of (\ref{eq:gradient-py}) w.r.t. $p(\mathbf{y})$
results in the following: 
\begin{equation}
\begin{aligned}\frac{\nabla_{\mathbf{y}}p(\mathbf{y})}{p(\mathbf{y})} & =\rho\int(\mathbf{h}-\mathbf{y})\frac{p(\mathbf{y}|\mathbf{h})p(\mathbf{h})}{p(\mathbf{y})}d\mathbf{h}\\
 & =\rho\int(\mathbf{h}-\mathbf{y})p(\mathbf{h}|\mathbf{y})d\mathbf{h}\\
 & =\rho\int\mathbf{h}p(\mathbf{h}|\mathbf{y})d\mathbf{h}-\rho\mathbf{y}\int p(\mathbf{h}|\mathbf{y})d\mathbf{h}\\
 & =\rho\left(\mathbf{\hat{h}}_{\text{MMSE}}-\mathbf{y}\right)\text{,}
\end{aligned}
\end{equation}
where the second equality holds owing to the Bayes\textquoteright{}
theorem, and the last equality holds due to (\ref{eq:posterior-mean})
and $\int p(\mathbf{h}|\mathbf{y})d\mathbf{h}=1$. By rearranging
the terms and plugging in $\frac{\nabla_{\mathbf{y}}p(\mathbf{y})}{p(\mathbf{y})}=\nabla_{\mathbf{y}}\log p(\mathbf{y})$,
we reach the foundation of the proposed algorithm: 
\begin{equation}
\hat{\mathbf{h}}_{\text{MMSE}}=\mathbf{y}+\frac{1}{\rho}\nabla_{\mathbf{y}}\log p(\mathbf{y}),\label{eq:Tweedie}
\end{equation}
where $\nabla_{\mathbf{y}}\log p(\mathbf{y})$ is called the \textit{Stein's
score function} in statistics \cite{2014Alain}. From (\ref{eq:Tweedie}),
we notice that the Bayes-optimal MMSE channel estimator can be achieved
\textit{solely based on the received pilot signals} $\mathbf{y}$,
without access to the prior distribution $p(\mathbf{h})$ or a supervised
dataset of the ground-truth channels $\mathbf{h}$, which are unavailable
in practice. With an efficient estimator of the score function $\nabla_{\mathbf{y}}\log p(\mathbf{y})$,
the Bayes-optimal MMSE channel estimator can be computed in a closed
form with extremely low complexity. In addition, since (\ref{eq:Tweedie})
holds regardless of the distribution of the HMIMO channel $\mathbf{h}$,
one can construct the MMSE estimator \textit{in arbitrary EM environments}
without any assumptions on the scatterers or the array geometry. 

It is noticed that two terms that should be obtained in (\ref{eq:Tweedie}),
i.e., the received SNR $\rho$ and the measurement score function
$S(\mathbf{y})$. In the following, we discuss on how to utilize machine
learning tools to obtain an accurate estimation of them based\textit{
solely} on the measurement $\mathbf{y}$. 

\subsection{Self-Supervised Learning of the Score Function}

We discuss how to get the score function $\nabla_{\mathbf{y}}\log p(\mathbf{y})$.
Given that a closed-form expression is intractable to acquire, we
instead aim to achieve a parameterized function with a neural network,
and discuss how to train it based on score matching. We first introduce
the denoising auto-encoder (DAE) \cite{2014Alain}, the core of the
training process, and explain how to utilize it to approximate the
score function. 

To obtain the score function $\nabla_{\mathbf{y}}\log p(\mathbf{y})$,
the measurement $\mathbf{y}$ is treated as the target signal that
the DAE should denoise. The general idea is to obtain the score function
$\nabla_{\mathbf{y}}\log p(\mathbf{y})$ based on its analytical relationship
with the DAE of $\mathbf{y}$, which will be established later in
\textbf{Theorem \ref{thm:The-optimal-DAE,}}. We first construct a
noisy version of the target signal $\mathbf{y}$ by \textit{manually}
adding some additive white Gaussian noise, $\sigma\mathbf{u}$, where
$\mathbf{u}\sim\mathcal{N}(\mathbf{0},\mathbf{I}_{2N})$ and $\text{\ensuremath{\sigma}}$
controls the noise level\footnote{Note that the extra noise is only added during the training process. },
and then train a DAE to \textit{denoise} the manually added noise.
The DAE, denoted by $R_{\bm{\theta}}(\cdot;\cdot)$, is trained by
the $\ell_{2}$-loss function, i.e., 
\begin{equation}
\mathcal{L}_{\text{DAE}}(\bm{\theta})=\mathbb{E}\|\mathbf{y}-R_{\bm{\theta}}(\mathbf{y};\sigma)\|^{2}.\label{eq:DAE-loss}
\end{equation}
 The theorem below explains the relationship between the score function
and the trained DAE. 
\begin{thm}[{Alain-Bengio \cite[Theorem 1]{2014Alain}}]
\label{thm:The-optimal-DAE,}The optimal DAE, $R_{\bm{\theta}^{*}}(\cdot;\cdot)$,
behaves asymptotically as
\begin{equation}
R_{\bm{\theta}^{*}}(\mathbf{y};\sigma)=\mathbf{y}+\sigma^{2}\nabla_{\mathbf{y}}\log p(\mathbf{y})+o(\sigma^{2}),\,\,\text{\text{as}}\,\,\sigma\rightarrow0.\label{eq:optimal DAE}
\end{equation}
\end{thm}
\begin{IEEEproof}
Please refer to \cite[Appendix A]{2014Alain}. 
\end{IEEEproof}
The above theorem indicates that, for a sufficiently small $\sigma$,
we can approximate the score function based on the DAE by $\nabla_{\mathbf{y}}\log p(\mathbf{y})\thickapprox\frac{R_{\bm{\theta}}(\mathbf{y};\sigma)-\mathbf{y}}{\sigma^{2}}$,
assuming that parameter of the DAE, $\bm{\theta}$, is near-optimal,
i.e., $\bm{\theta}\thickapprox\bm{\theta}^{*}$. Nevertheless, the
approximation can be numerically unstable as the denominator, $\text{\ensuremath{\sigma^{2}}}$,
is close to zero. To alleviate the problem, we improve the structure
of the DAE and rescale the original loss function. 

First, we consider a residual form of the DAE with a scaling factor.
Specifically, let $R_{\bm{\theta}}(\mathbf{y};\sigma)=\sigma^{2}S_{\bm{\theta}}(\mathbf{y};\sigma)+\mathbf{y}$.
Plugging it into (\ref{eq:optimal DAE}), the score function is approximately
equal to 
\begin{equation}
\nabla_{\mathbf{y}}\log p(\mathbf{y})\thickapprox\frac{(\sigma^{2}S_{\bm{\theta}}(\mathbf{y};\sigma)+\mathbf{y})-\mathbf{y}}{\sigma^{2}}=S_{\bm{\theta}}(\mathbf{y};\sigma),\label{eq:approximate-AR-DAE}
\end{equation}
when $\sigma\rightarrow0$ holds. This reparameterization enables
$S_{\bm{\theta}}(\mathbf{y};\sigma)$ to approximate the score function
directly, thereby circumventing the need for division that may lead
to numerical instability. Also, the residual link significantly enhances
the capability of the DAE, since it can easily learn an identity mapping
\cite{2016He}. 

Second, since the variance $\sigma^{2}$ of the manually added noise
is small, the gradient of the DAE loss function (\ref{eq:DAE-loss})
can easily vanish to zero and may lead to difficulties in training.
Hence, we rescale the loss function by a factor of $\frac{1}{\sigma}$
to safeguard the vanishing gradient problem, i.e., 
\begin{equation}
\text{\ensuremath{\mathcal{L}_{\text{DAE}}(\bm{\theta})=\mathbb{E}\|\mathbf{u}+\sigma S_{\bm{\theta}}(\mathbf{y};\sigma)\|^{2}}},\label{eq:improved-DAE-loss}
\end{equation}
where (\ref{eq:approximate-AR-DAE}) is plugged into the loss function. 

We are interested in the region where $\sigma$ is sufficiently close
to zero, in which case $S_{\bm{\theta}}(\mathbf{y};0)$ can be deemed
to be equal to the score function $\nabla_{\mathbf{y}}\log p(\mathbf{y})$
according to (\ref{eq:approximate-AR-DAE}). Nevertheless, directly
training the network using a very small $\sigma$ is difficult since
the SNR of the gradient signal decreases in a linear rate $\mathcal{O}(\sigma)$
with respect to $\sigma$, which introduces difficulty for the stochastic
gradient descent \cite{2020Lim}. To exploit the asymptotic optimality
of the score function approximation when $\sigma\rightarrow0$, we
propose to simultaneously train the network $S_{\bm{\theta}}(\mathbf{y};\sigma)$
with varying $\sigma$ values, to handle various $\sigma$ levels
and then naturally generalize to the desired region, i.e., $S_{\bm{\theta}}(\mathbf{y};0)$.
To achieve the goal, we control the manually added noise by letting
$\sigma$ follow a zero-mean Gaussian distribution $\sigma\sim\mathcal{N}(0,\text{\ensuremath{\xi^{2}}})$
and gradually anneal $\xi\in[\sigma_{\text{min}},\sigma_{\text{max}}]$
from a large value $\sigma_{\text{max}}$ to a small one $\sigma_{\text{min}}\thickapprox0$
in each iteration. That is, we \textit{condition} $S_{\bm{\theta}}(\mathbf{y};\sigma)$
on the manually added noise level $\sigma$ during training. 

The proposed algorithm is shown in \textbf{Algorithm 1}. The DAE is
trained using stochastic gradient descent for $Q$ epochs. In each
epoch, we draw a random vector $\mathbf{u}$ and anneal $\xi$ in
$\sigma\sim\mathcal{N}(0,\text{\ensuremath{\xi^{2}}})$ to control
the extra noise level according to the current number of iterations
$q$. Then, the DAE loss function $\mathcal{L}_{\text{DAE}}$ in (\ref{eq:improved-DAE-loss})
is minimized by stochastic optimization. Note that in the training
process, nothing but a dataset of the received pilot signals $\mathbf{y}$
is necessary, which is readily available in practice. In the inference
stage, one can apply formula (\ref{eq:Tweedie}) to compute the score-based
MMSE estimator, in which the score function can be approximated by
using $S_{\bm{\theta}}(\mathbf{y};0)$, i.e., setting $\sigma$ as
zero, and the received SNR $\hat{\rho}$ can be estimated by the PCA-based
algorithm in the next subsection. 

For the neural architecture of the DAE $S_{\bm{\theta}}(\cdot;\cdot)$,
we adopt a simplified UNet architecture \cite{2015Ronneberger}. Note
that depending on the complexity budget, many other prevailing neural
architectures could also be applied \cite{2023Yu-AI}. Other details
of the training process are deferred to Section \ref{sec:Simulation-Results}.
\floatplacement{algorithm}{t}
\begin{algorithm} 
\caption{Training and inference of the proposed algorithm}
\begin{algorithmic}[1]

\STATE \emph{/* The offline training stage */}
\STATE {\bf Input:} Learning rate $\gamma$, maximum extra noise level $\sigma_{\text{max}}$, minimum extra noise level $\sigma_{\text{min}}$, number of epochs $Q$, a dataset of received pilots $\{\mathbf{y}_i\}_{i=1}^M$ \\
\STATE {\bf Output:} Trained DAE parameters $\bm{\theta}$ \\
\STATE {\bf for} $q=1:Q$ {\bf do}
\STATE\hspace{\algorithmicindent} Draw $\mathbf{u} \sim \mathcal{N}(\mathbf{0},\mathbf{I}_{2N})$
\STATE\hspace{\algorithmicindent} Set the extra noise level with $\xi \gets \frac{Q-q}{Q}\sigma_{\text{min}}+\frac{q}{Q}\sigma_{\text{max}}$
\STATE\hspace{\algorithmicindent} Compute the loss function $\mathcal{L}_\text{DAE}$ as in (18)
\STATE\hspace{\algorithmicindent} Update NN parameters as $\bm{\theta} \gets \bm{\theta}-\gamma \nabla_{\bm{\theta}}\mathcal{L}_\text{DAE}$
\STATE {\bf return} $\bm{\theta}$  \\
\vspace*{\baselineskip}

\STATE \emph{/* The online inference stage */}
\STATE {\bf Input:} Received pilots $\mathbf{y}$, trained DAE parameters $\bm{\theta}$, size of the sliding window $\sqrt{d}\times\sqrt{d}$
\STATE {\bf Initialize:} $\nabla_\mathbf{y} \log p(\mathbf{y}) \gets S_{\bm{\theta}}(\mathbf{y};0)$
\STATE Utilize the PCA-based algorithm \cite{2023Yu} to estimate SNR $\hat{\rho}$
\STATE Compute the estimated channel $\mathbf{\hat{h}}_{\text{MMSE}}$ as in (14)
\STATE {\bf return} $\mathbf{\hat{h}}_{\text{MMSE}}$  \\
\end{algorithmic}
\end{algorithm}

\subsection{PCA-Based Received SNR Estimation}

We propose a low-complexity PCA-based algorithm to estimate the received
SNR $\rho$ in (\ref{eq:Tweedie}) based on \textit{a single instance}
of the pilot signals $\mathbf{y}$. Before further discussion, we
stress that the received SNR estimation is executed only at the inference
stage, not at the training stage, as shown in $\textbf{Algorithm 1}$. 

The basic idea behind the PCA-based algorithm is the low-rankness
of the spatial correlation matrix $\mathbf{R}$ of HMIMO due to the
dense deployment of the antenna elements. Specifically, for isotropic
scattering environments, the rank of the correlation matrix $\mathbf{R}_{\text{iso}}$
is approximately $\text{rank}(\mathbf{R}_{\text{iso}})\thickapprox\pi Nd_{a}^{2}/\lambda_{c}^{2}$
\cite{2022Pizzo}. It decreases with the shrink of antenna spacing
and the increase of the carrier frequency. For example, when $d_{a}=\lambda_{c}/4$,
around $80\%$ of the eigenvalues of $\mathbf{R}_{\text{iso}}$ shrinks
towards zero. The rank deficiency of $\mathbf{R}$ tends to be even
more prominent in the case of non-isotropic scattering environments
\cite{2023An}. 

Similar to Fig. 2 in our previous work \cite{2023Yu}, we decompose
multiple virtual subarray channels (VSCs) from the HMIMO channel by
a sliding window. Specifically, we reshape the real-valued HMIMO channel
$\mathbf{h}\in\mathbb{R}^{2N\times1}$ into a tensor form $\mathbf{H}\in\mathbb{R}^{\sqrt{N}\times\sqrt{N}\times2}$.
We then decompose $\mathbf{H}$ into $s=(\sqrt{N}-\sqrt{d}+1)^{2}$
VSC tensors using a sliding window of size $\mathbb{R}^{\sqrt{d}\times\sqrt{d}\times2}$,
and then reshape them back into the vector form to obtain a set of
VSCs, denoted by $\{\mathbf{h}_{t}\in\mathbb{R}^{2d\times1}\}_{t=1}^{s}$.
Similarly, the received pilot signals and the noise could also be
decomposed as $\{\mathbf{y}_{t}\in\mathbb{R}^{2d\times1}\}_{t=1}^{s}$
and $\{\mathbf{n}_{t}\in\mathbb{R}^{2d\times1}\}_{t=1}^{s}$, and
should satisfy 
\begin{equation}
\mathbf{y}_{t}=\sqrt{\rho}\mathbf{h}_{t}+\mathbf{n}_{t}.
\end{equation}

\begin{figure}[t]
\centering{}\includegraphics[width=6.3cm]{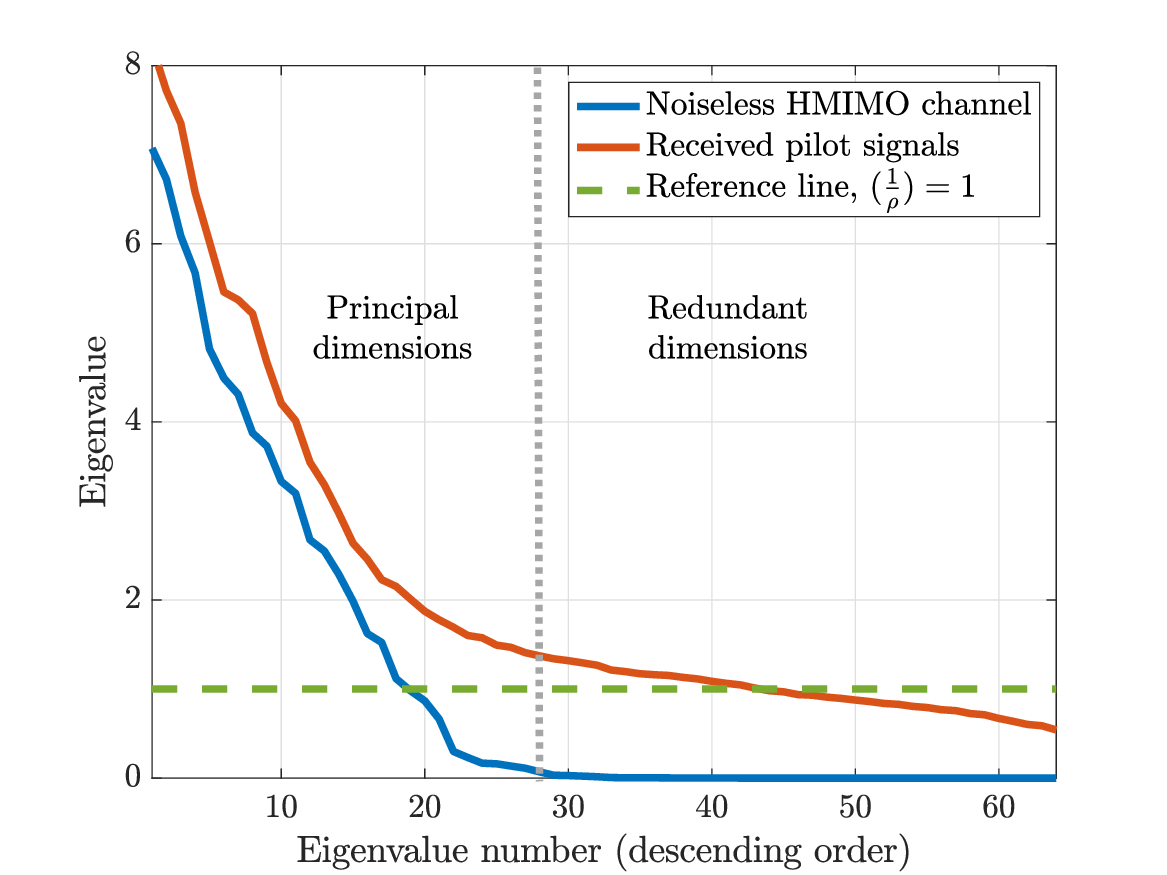}\caption{Eigenvalues of the covariance matrices in descending order from an
HMIMO channel in isotropic scattering environments and their corresponding
received pilots, when the received SNR is 0 dB. \label{fig:Eigenvalues-of-the}}
\end{figure}

Due to the low-rankness of the spatial correlation matrix $\mathbf{R}$,
the decomposed VSCs $\mathbf{h}_{t}$ should also lie in a low-dimensional
subspace. In Fig. \ref{fig:Eigenvalues-of-the}, we plot the eigenvalues
of the covariance matrices of the decomposed VSCs $\{\mathbf{h}_{t}\in\mathbb{R}^{2d\times1}\}_{t=1}^{s}$
from an HMIMO channel in isotropic scattering environments and their
corresponding received pilots $\{\mathbf{y}_{t}\in\mathbb{R}^{2d\times1}\}_{t=1}^{s}$,
respectively, with a reference line marking the inverse of the received
SNR $\frac{1}{\rho}$. According to the figure, the eigenvalues of
the covariance of $\mathbf{h}_{t}$ quickly shrinks to zero with about
30 principal dimensions. The zero eigenvalues correspond to the redundant
dimensions. In contrast, we observe that the eigenvalues of the covariance
of $\mathbf{y}_{t}$ are concentrated around $\frac{1}{\rho}$ in
the redundant dimensions. This example suggests that it is indeed
possible to estimate the received SNR based on the redundant eigenvalues.
Rigorously, these eigenvalues could be proved to follow a Gaussian
distribution $\mathcal{N}(\frac{1}{\rho},\frac{2}{s\rho^{2}})$ \cite{2023Yu}.
Plus, the principal and redundant eigenvalues could be separated by
an iterative process. Hence, we could leverage \cite[Algorithm 1]{2023Yu}
to accurately estimate the received SNR. Detailed setups are discussed
in Section \ref{sec:Simulation-Results}. 

\subsection{Complexity Analysis}

The inference complexity of the proposed algorithm consists of the
computation of $S_{\bm{\theta}}(\mathbf{y};0)$ and the PCA-based
estimation of $\rho$. The former depends upon the specific neural
architecture of $S_{\bm{\theta}}(\cdot,\cdot)$, and costs a constant
complexity, denoted by $p$, once the network is trained. The computational
complexity for the latter, as analyzed in \cite{2023Yu}, is given
by $\text{\ensuremath{\mathcal{O}}}(Nd^{2}+d^{3})$, where $d$, the
size of the sliding window, is usually quite small. Hence, the overall
complexity is $\text{\ensuremath{\mathcal{O}}}(Nd^{2}+d^{3}+p)$,
which scales only \textit{linearly} with respect to the number of
antennas $N$. 

In practice, the fluctuation of the received SNR may not be frequent.
In such a case, it is not a necessity to estimate the received SNR
for every instance of the received pilot signals $\mathbf{y}$. Therefore,
the actual complexity of the proposed score-based algorithm is within
the range of $\mathcal{O}(p)$ and $\text{\ensuremath{\mathcal{O}}}(Nd^{2}+d^{3}+p)$,
which is extremely efficient given near-optimal performance\footnote{Most previous works on channel estimation assume that the received
SNR is perfectly known. In this case, the complexity reduces to $\mathcal{O}(p)$. }. By sharp contrast, the (oracle) MMSE estimator requires time-consuming
matrix inversion, which is as complex as $\mathcal{O}(N^{3})$. In
Section \ref{sec:Simulation-Results}, we provide a running time complexity
to offer a straightforward comparison. 

\section{Simulation Results\label{sec:Simulation-Results}}

\subsection{Simulation Setups}

We consider a typical HMIMO system setup with $N=1024$ and $d_{a}=\lambda_{c}/4$.
In the training stage, the hyper-parameters of the proposed score-based
algorithm are chosen as $\gamma=0.001$, $\sigma^{\text{min}}=0.001$,
$\sigma^{\text{max}}=0.1$, $Q=100$, and $M=10\text{,}000$. Also,
the learning rate $\gamma$ is decayed by half after every 25 epochs.
In the inference stage, the size of the sliding window is set as $7\times7$.
The performance discussed below is all averaged over a held-out dataset
consisting of $L=10,000$ testing samples. 
\begin{table}[t]
\begin{centering}
\caption{Performance of the received SNR estimation \label{tab:Performance-of-the-SNR}}
\par\end{centering}
\centering{}%
\begin{tabular}{c|c|>{\centering}p{0.9cm}|>{\centering}p{0.7cm}|>{\centering}p{0.7cm}}
\hline 
$\frac{1}{\rho}$ / (SNR) & Method & Bias & Std & RMSE\tabularnewline
\hline 
\multirow{3}{*}{1.0000 (0 dB)} & Oracle & 0.0004 & 0.0160 & 0.0160\tabularnewline
\cline{2-5} \cline{3-5} \cline{4-5} \cline{5-5} 
 & \textbf{Proposed} & 0.0080 & 0.0371 & 0.0380\tabularnewline
\cline{2-5} \cline{3-5} \cline{4-5} \cline{5-5} 
 & Sparsity & 0.1769 & 0.0277 & 0.1790\tabularnewline
\hline 
\multirow{3}{*}{0.1778 (15 dB)} & Oracle & <0.0001 & 0.0028 & 0.0028\tabularnewline
\cline{2-5} \cline{3-5} \cline{4-5} \cline{5-5} 
 & \textbf{Proposed} & 0.0031 & 0.0064 & 0.0071\tabularnewline
\cline{2-5} \cline{3-5} \cline{4-5} \cline{5-5} 
 & Sparsity & 0.1530 & 0.0163 & 0.1539\tabularnewline
\hline 
\end{tabular}
\end{table}

For isotropic scattering, the covariance $\mathbf{R}_{\text{iso}}$
is given via (\ref{eq:isotropic-covariance}). In the non-isotropic
case, we follow \cite{2023An} and also construct the covariance $\mathbf{R}$
by truncating the leading $\nicefrac{\text{rank}(\mathbf{R}_{\text{iso}})}{8}$
eigenvalues of $\mathbf{R}_{\text{iso}}$. Mathematically, the non-isotropic
covariance could be expressed as $\mathbf{R}=\mathbf{V}\boldsymbol{\Lambda}_{\text{trunc}}\mathbf{V}^{H}$,
where $\mathbf{V}$ is a matrix consisting of the eigenvectors of
$\mathbf{R}_{\text{iso}}$, while $\boldsymbol{\Lambda}_{\text{trunc}}$
is a diagonal matrix that contains the eigenvalues of $\mathbf{R}_{\text{iso}}$
arranged in descending order. To truncate the matrix, we select only
the first $\nicefrac{\text{rank}(\mathbf{R}_{\text{iso}})}{8}$ eigenvalues
and eigenvectors of $\mathbf{R}_{\text{iso}}$. 

\subsection{Accuracy of Received SNR Estimation}

The accuracy of the received SNR estimation can influence the performance
of the proposed score-based estimator. Hence, different from previous
works that assume perfect knowledge of the received SNR, we propose
a practical PCA-based means to estimate it and examine its performance.
In Table \ref{tab:Performance-of-the-SNR}, we list the estimation
accuracy under different SNRs. We present the performance in terms
of the inverse of the SNR, i.e., $\frac{1}{\rho}$, since it is used
in (\ref{eq:Tweedie}). The bias, the standard deviation (std), and
the root MSE (RMSE) are given by $\mathbb{E}[|\frac{1}{\rho}-\mathbb{E}[(\frac{1}{\hat{\rho}})]|]$,
$\sqrt{\mathbb{E}[((\frac{1}{\hat{\rho}})-\mathbb{E}[(\frac{1}{\hat{\rho}})])^{2}]}$,
and $\sqrt{\mathbb{E}[(\frac{1}{\rho}-(\frac{1}{\hat{\rho}}))^{2}]}$,
respectively. The estimator's accuracy and robustness are reflected
with the bias and std, while the RMSE provides an assessment of its
overall performance. From Table \ref{tab:Performance-of-the-SNR},
we observe that the performance of the proposed PCA-based method is
both highly accurate and robust, and outperforms the sparsity-based
median absolute deviation (MAD) estimator in \cite{2023Gallyas-Sanhueza},
as HMIMO channels are not exactly sparse. Also, the performance is
close to the oracle bound which assumes perfect knowledge of the channel
$\mathbf{h}$, and estimate $\rho$ directly from $\mathbf{y-\mathbf{h}}$
\cite{2023Yu}. Later, in Section \ref{subsec:Robustness-to-SNR},
we will illustrate that the proposed score-based channel estimator
is robust to SNR estimation errors. 
\begin{figure*}[t]
\centering{}\subfloat[\label{fig:NMSE-iso}]{\centering{}\includegraphics[width=6cm]{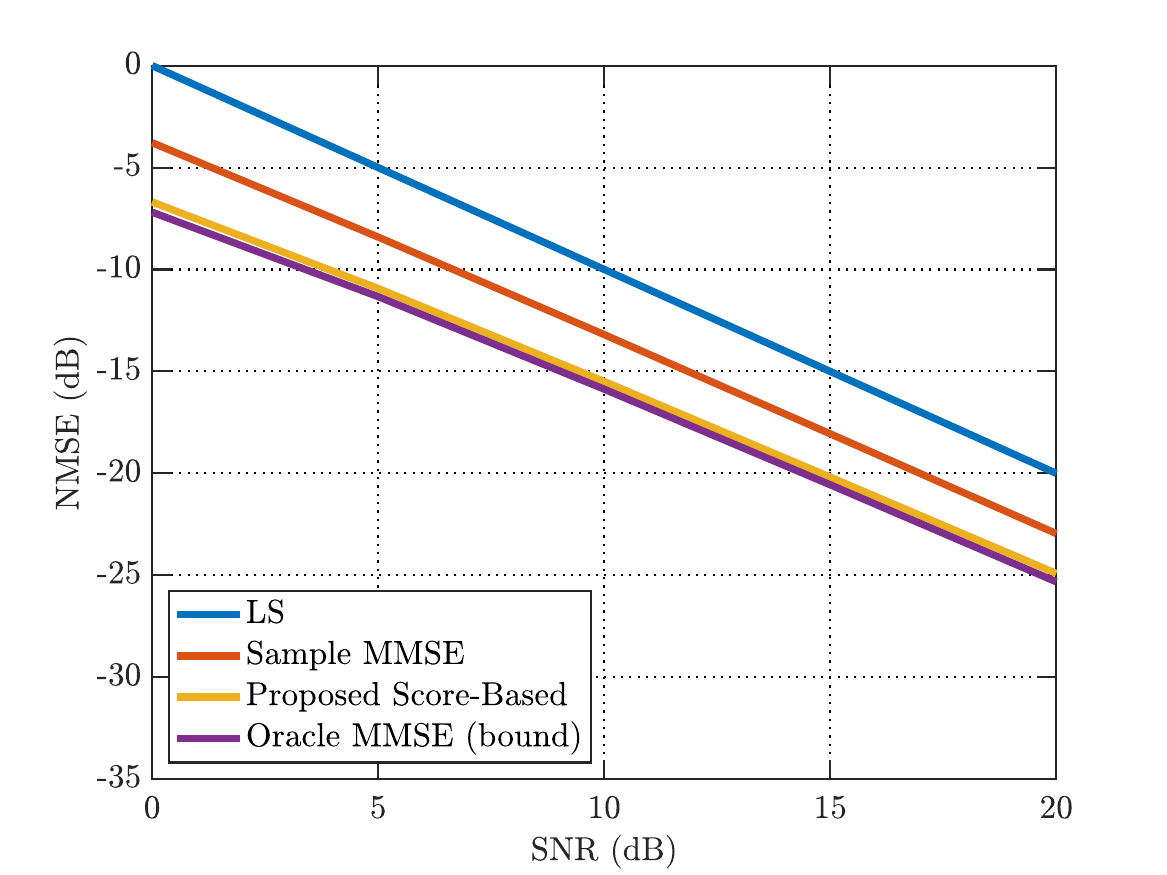}}\subfloat[\label{fig:NMSE-non-iso}]{\centering{}\includegraphics[width=6cm]{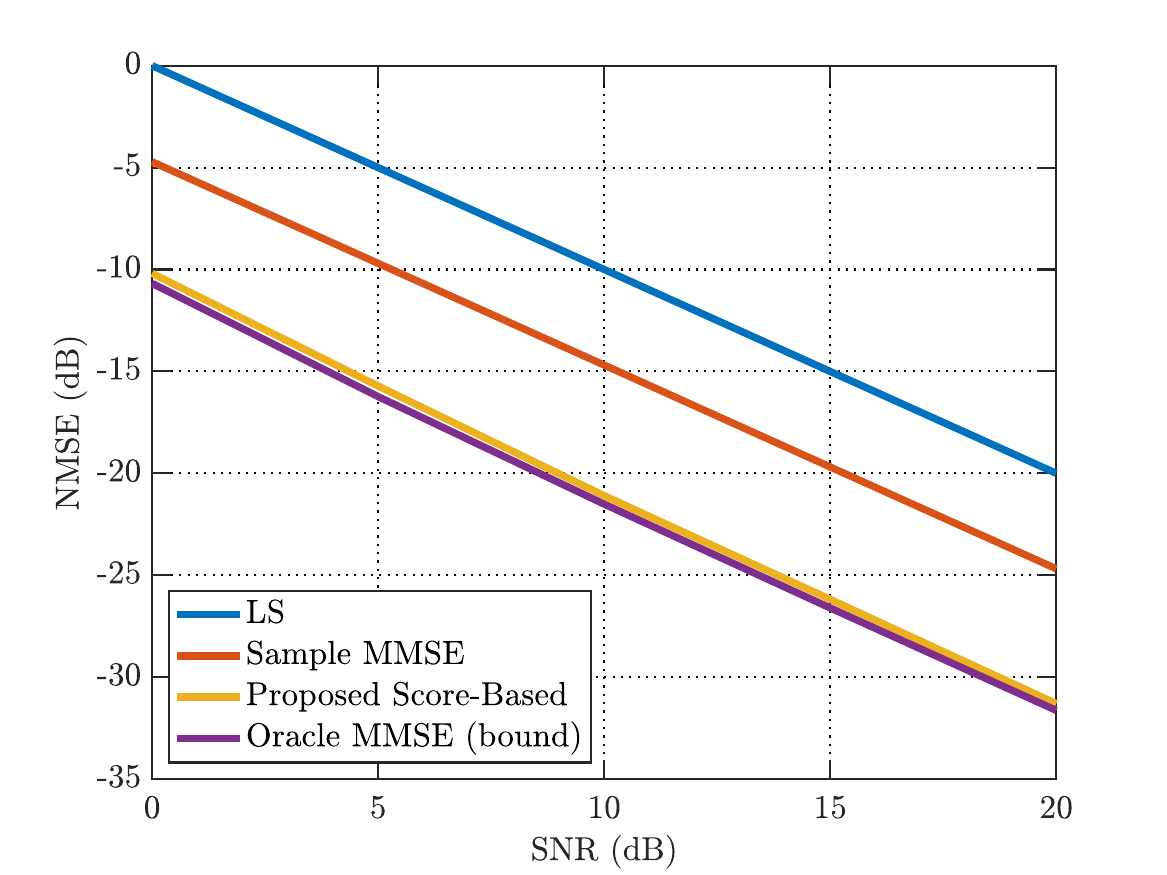}}\subfloat[\label{fig:Robustness}]{\centering{}\includegraphics[width=6cm]{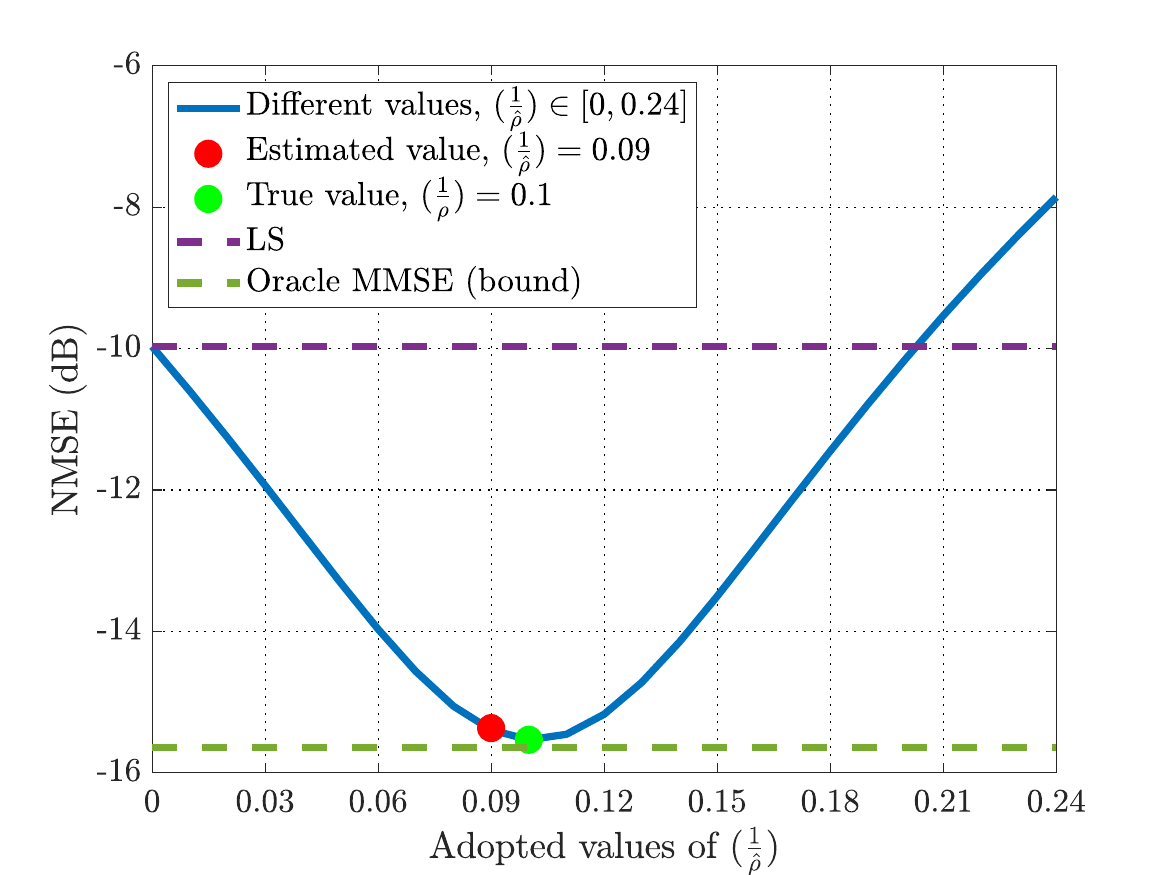}}\caption{Simulation results. (a) NMSE versus SNR in \textit{isotropic} scattering.
(b) NMSE versus SNR in \textit{non-isotropic} scattering. (c) The
influence of the accuracy of the received SNR estimation on the NMSE
performance, when the received SNR is $\rho=10$ dB. \label{fig:Simulation-results}}
\end{figure*}

\subsection{Normalized MSE (NMSE) in Isotropic and Non-Isotropic Scattering Environments}

We compare the proposed score-based estimator with three benchmarks,
including LS, sample MMSE, and oracle MMSE. Here, oracle MMSE refers
to the MMSE estimator with \textit{perfect knowledge} of both the
covariance $\mathbf{R}$ and the received SNR $\rho$, which is the
Bayesian performance bound, given by 
\begin{equation}
\mathbf{\hat{h}}_{\text{oracle-MMSE}}=\sqrt{\rho}\mathbf{R}(\rho\mathbf{R}+\mathbf{I})^{-1}\mathbf{y}.\label{eq:oracle-MMSE}
\end{equation}
This is difficult, if not impossible, to acquire in practice since
$\mathbf{R}$ contains $N^{2}$ entries and is prohibitive to estimate
when $N$ is particularly large in an HMIMO system. The sample MMSE
method utilizes the same equation as (\ref{eq:oracle-MMSE}), but
replaces the true covariance $\mathbf{R}$ with an estimated one $\mathbf{R}_{\text{sample}}$
based upon the testing samples, given by $\mathbf{R}_{\text{sample}}\triangleq\frac{1}{L}\sum_{l=1}^{L}\mathbf{y}_{l}\mathbf{y}_{l}^{H}-\frac{1}{\rho}\mathbf{I}$
\cite{2023Damico}, where $L$ is the number of testing samples and
$\mathbf{y}_{l}$ denotes the $l$-th sample of the received pilot
signals in the testing dataset. We also utilize the perfect received
SNR $\rho$ in sample MMSE. 

In Fig. \ref{fig:Simulation-results}(a), we present the NMSE as a
function of the SNR $\rho$ in \textit{isotropic} scattering environments.
It is illustrated that the proposed score-based algorithm significantly
outperforms the LS and the sample MMSE estimator, and achieves almost
the same NMSE as the oracle MMSE bound at every SNR level. Note that
the oracle MMSE method utilizes the true covariance and received SNR,
but the proposed method requires neither. 

In Fig. \ref{fig:Simulation-results}(b), we compare the NMSE in \textit{non-isotropic}
environments, in which the rank of the covariance matrix $\mathbf{R}$
is further reduced. As a result, the performance gap between the LS
and the oracle MMSE estimator is enlarged as the rank deficiency becomes
more significant. Nevertheless, similar to the isotropic case, the
proposed score-based estimator exhibits almost the same performance
as the oracle bound and significantly outperforms the sample MMSE
method, illustrating its effectiveness in different scattering environments. 

\subsection{Robustness to SNR Estimation Errors\label{subsec:Robustness-to-SNR}}

In Fig. \ref{fig:Simulation-results}(c), we provide discussions on
how the accuracy for the received SNR estimation will influence the
performance of the proposed score-based estimator. In the simulations,
the true value of the received SNR is set as 10 dB, i.e., $\frac{1}{\rho}=0.1$.
We vary the adopted values of $(\frac{1}{\hat{\rho}})$ in (\ref{eq:Tweedie})
within $(\frac{1}{\hat{\rho}})\in[0,0.24]$, and plot the corresponding
NMSE performance curve in blue. Particularly, we utilize red and green
dots to denote the NMSE achieved with the estimated and true SNR values,
respectively. The LS and the oracle MMSE algorithms are presented
as the performance upper and lower bounds. It is observed that even
when an inexact received SNR is adopted, the performance of the score-based
algorithm is still quite robust and significantly outperforms the
LS method. Also, the estimated received SNR by the PCA-based method
is accurate enough to offer a near-optimal performance, even in unknown
EM environments. 

\subsection{Running Time Complexity}

We introduce the CPU running time of the proposed algorithm. For the
considered setups, the proposed score-based estimator takes as low
as 3 ms on an Intel Core i7-9750H CPU, which is much shorter than
that of the oracle MMSE method involving high-dimensional matrix inverse
(requiring around 70 ms). The high efficiency and the Bayes-optimal
performance in unknown EM environments thus make the proposed score-based
algorithm an ideal candidate in practice. 

\section{Conclusion and Future Directions}

In this paper, we studied channel estimation for the HMIMO systems,
and proposed a score-based MMSE channel estimator that can achieve
Bayes-optimal performance with an extremely low complexity. Particularly,
the proposed algorithm is trained solely based on the received pilots,
without requiring any kind of priors or supervised datasets that are
prohibitive to collect in practice. This enables it to work in arbitrary
and unknown EM environments that may appear in real-world deployment.
As a future direction, it is interesting to extend to the proposed
framework to compressed sensing-based setups. Furthermore, since the
proposed algorithm is self-supervised, it is promising to investigate
the possibility of online learning and adaptation. 

\bibliographystyle{IEEEtran}
\bibliography{references_ideas}

\end{document}